\documentclass[a4paper,11pt]{article}
\usepackage{pos}

\title{The Compton Spectrometer and Imager}
 \ShortTitle{The Compton Spectrometer and Imager}

\author*[a]{John A. Tomsick}
\author[b,a]{Steven E. Boggs}
\author[a]{Andreas Zoglauer}
\author[c]{Dieter Hartmann}
\author[c]{Marco Ajello}
\author[d]{Eric Burns}
\author[e]{Chris Fryer}
\author[f,w]{Chris Karwin}
\author[f]{Carolyn Kierans}
\author[a]{Alexander Lowell}
\author[g]{Julien Malzac}
\author[b]{Jarred Roberts}
\author[a]{Pascal Saint-Hilaire}
\author[f]{Albert Shih}
\author[h]{Thomas Siegert}
\author[i]{Clio Sleator}
\author[j]{Tadayuki Takahashi}
\author[k]{Fabrizio Tavecchio}
\author[i]{Eric Wulf}
\author[a]{Jacqueline Beechert}
\author[a]{Hannah Gulick}
\author[a]{Alyson Joens}
\author[a]{Hadar Lazar}
\author[f,l]{Eliza Neights}
\author[a]{Juan Carlos Martinez Oliveros}
\author[j]{Shigeki Matsumoto}
\author[j]{Tom Melia}
\author[h]{Hiroki Yoneda}
\author[m]{Mark Amman}
\author[c]{Dhruv Bal}
\author[g]{Peter von Ballmoos}
\author[c]{Hugh Bates}
\author[n]{Markus B\"ottcher}
\author[o]{Andrea Bulgarelli}
\author[p]{Elisabetta Cavazzuti}
\author[q]{Hsiang-Kuang Chang}
\author[a]{Claire Chen}
\author[q]{Che-Yen Chu}
\author[o]{Alex Ciabattoni}
\author[p]{Luigi Costamante}
\author[n]{Lente Dreyer}
\author[o]{Valentina Fioretti}
\author[p]{Francesco Fenu}
\author[r]{Savitri Gallego}
\author[k]{Giancarlo Ghirlanda}
\author[i]{Eric Grove}
\author[q]{Chien-You Huang}
\author[g]{Pierre Jean}
\author[c]{Nikita Khatiya}
\author[g]{J\"urgen Kn\"odlseder}
\author[s]{Martin Krause}
\author[c]{Mark Leising}
\author[f,w]{Tiffany R. Lewis}
\author[r]{Jan Peter Lommler}
\author[t]{Lea Marcotulli}
\author[f,u]{Israel Martinez-Castellanos}
\author[h]{Saurabh Mittal}
\author[d]{Michela Negro}
\author[a]{Samer Al Nussirat}
\author[v]{Kazuhiro Nakazawa}
\author[r]{Uwe Oberlack}
\author[d]{David Palmore}
\author[o]{Gabriele Panebianco}
\author[o]{Nicolo Parmiggiani}
\author[f]{Tyler Parsotan}
\author[b]{Sean N. Pike}
\author[a]{Field Rogers}
\author[n]{Hester Schutte}
\author[c]{Yong Sheng}
\author[f]{Alan P. Smale}
\author[i]{Jacob Smith}
\author[d]{Aaron Trigg}
\author[f]{Tonia Venters}
\author[j]{Yu Watanabe}
\author[f,w]{Haocheng Zhang}

\affiliation[a]{Space Sciences Laboratory, 7 Gauss Way, University of California, Berkeley CA 94720-7450, USA}

\affiliation[b]{Department of Astronomy \& Astrophysics, UC San Diego, 9500 Gilman Drive, La Jolla CA 92093, USA}

\affiliation[c]{Clemson University, South Carolina, USA}

\affiliation[d]{Louisiana State University, Baton Rouge, LA 70803, USA}

\affiliation[e]{Los Alamos National Laboratory, New Mexico, USA}

\affiliation[f]{NASA Goddard Space Flight Center, 8800 Greenbelt Road, Greenbelt, MD 20771, USA}

\affiliation[g]{Institut de Recherche en Astrophysique et Planetologie, 9, avenue du Colonel Roche BP 44346 31028 Toulouse Cedex 4, France}

\affiliation[h]{Institut f{\"u}r Theoretische Physik und Astrophysik, Universit{\"a}t W{\"u}rzburg, Campus Hubland Nord, Emil-Fischer-Str. 31, 97074, W{\"u}rzburg, Germany}

\affiliation[i]{U.S. Naval Research Laboratory, 4555 Overlook Ave., SW Washington, DC 20375, USA}

\affiliation[j]{Kavli Institute for the Physics and Mathematics of the Universe, The University of Tokyo, Japan}

\affiliation[k]{Istituto Nazionale di Astrofisica, Merate, Italy}

\affiliation[l]{George Washington University, USA}

\affiliation[m]{Independent, USA}

\affiliation[n]{Centre for Space Research, North-West University, Potchefstroom 2520, South Africa}

\affiliation[o]{Istituto Nazionale di Astrofisica, Bologna, Italy}

\affiliation[p]{Italian Space Agency, Italy}

\affiliation[q]{Institute of Astronomy, National Tsing Hua University, Guangfu Rd., Hsinchu City, 300044, Taiwan}

\affiliation[r]{Institut f{\"u}r Physik \& Exzellenzcluster PRISMA\textsuperscript{+}, Johannes Gutenberg-Universit{\"a}t Mainz, 55099 Mainz, Germany}

\affiliation[s]{Centre for Astrophysics Research, Department of Physics, Astronomy and Mathematics, University of Hertfordshire, College Lane, Hatfield AL10 9AB, UK}

\affiliation[t]{Yale University, USA}

\affiliation[u]{University of Maryland, USA}

\affiliation[v]{Nagoya University, Japan}

\affiliation[w]{NASA Postdoctoral Program Fellow}

\emailAdd{jtomsick@berkeley.edu}

\abstract{{\bf Abstract:} The Compton Spectrometer and Imager (COSI) is a NASA Small Explorer (SMEX) satellite mission in development with a planned launch in 2027. COSI is a wide-field gamma-ray telescope designed to survey the entire sky at 0.2-5\,MeV.  It provides imaging, spectroscopy, and polarimetry of astrophysical sources, and its germanium detectors provide excellent energy resolution for emission line measurements.  Science goals for COSI include studies of 0.511 MeV emission from antimatter annihilation in the Galaxy, mapping radioactive elements from nucleosynthesis, determining emission mechanisms and source geometries with polarization measurements, and detecting and localizing multimessenger sources.  The instantaneous field of view for the germanium detectors is $>$25\% of the sky, and they are surrounded on the sides and bottom by active shields, providing background rejection as well as allowing for detection of gamma-ray bursts and other gamma-ray flares over most of the sky.  In the following, we provide an overview of the COSI mission, including the science, the technical design, and the project status.}

\ConferenceLogo{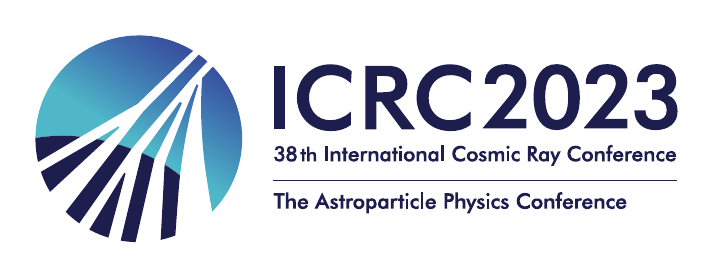}

\FullConference{%
38th International Cosmic Ray Conference (ICRC2023)\\
  26 July - 3 August, 2023\\
  Nagoya, Japan}

\begin{document}
\maketitle

\section{Overview}
\vspace{-0.3cm}
The low-energy part of the gamma-ray regime -- the "MeV bandpass" -- has not been well-explored due to observational challenges, such as high background levels.  However, this is a part of the electromagnetic spectrum where we know that there are signals of great scientific interest, including nuclear lines from radioactive elements and the electron-positron annihilation line at 0.511\,MeV.  In addition, accreting black holes produce emission at these energies, and this is also a key regime for multimessenger astrophysics (MMA).

Future observations made by the COSI satellite mission will address all of these topics, and its key scientific goals are related to positrons, nuclear lines, polarimetry of accreting black holes, and MMA \cite{tomsick22}.  COSI is a Compton telescope, and the capabilities that make it unique are its combination of excellent energy resolution, which it gets from cryogenically-cooled germanium detectors, and its large field of view.  Thus, COSI is optimized to make all-Galaxy and all-sky emission line images in the MeV bandpass, which will advance our understanding of creation and destruction of matter in our Galaxy.

Here, we provide an update on the COSI mission, focusing on developments since our previous ICRC presentation in July 2021 \cite{tomsick22}.  COSI was selected by NASA to move to Phase B in October 2021.  This was followed by maturing of the requirements and the payload design, and COSI completed a combined System Requirements Review and Mission Definition Review during the first calendar quarter of 2023.  

In this proceedings paper, we describe the COSI mission concept and design in Section 2.  A summary of how the COSI design enables the key science goals is provided in Section 3.  Examples of the extended portfolio of science that will be enabled by COSI are given in Section 4.

\section{COSI mission concept and design}
\vspace{-0.3cm}
COSI operates as a Compton telescope when multiple photon interactions are detected in the instrument.  This occurs when a gamma-ray enters the instrument and undergoes one or more Compton scattering events before the interactions end in a final photoabsorption event.  When the full energy of a gamma-ray is deposited in the germanium detectors, there are two quantities that are obtained: the energy of the gamma-ray is determined by adding up the individual interaction energies; and the Compton angle measured from an axis defined by the first and second interactions defines an event circle on the sky (\cite{zoglauer21} and references therein).  Thus, it is essential that the detectors provide precise energy measurements and 3-dimensional interaction positions.

\begin{figure}[t]
\begin{center}
\includegraphics[height=3.5in,angle=0]{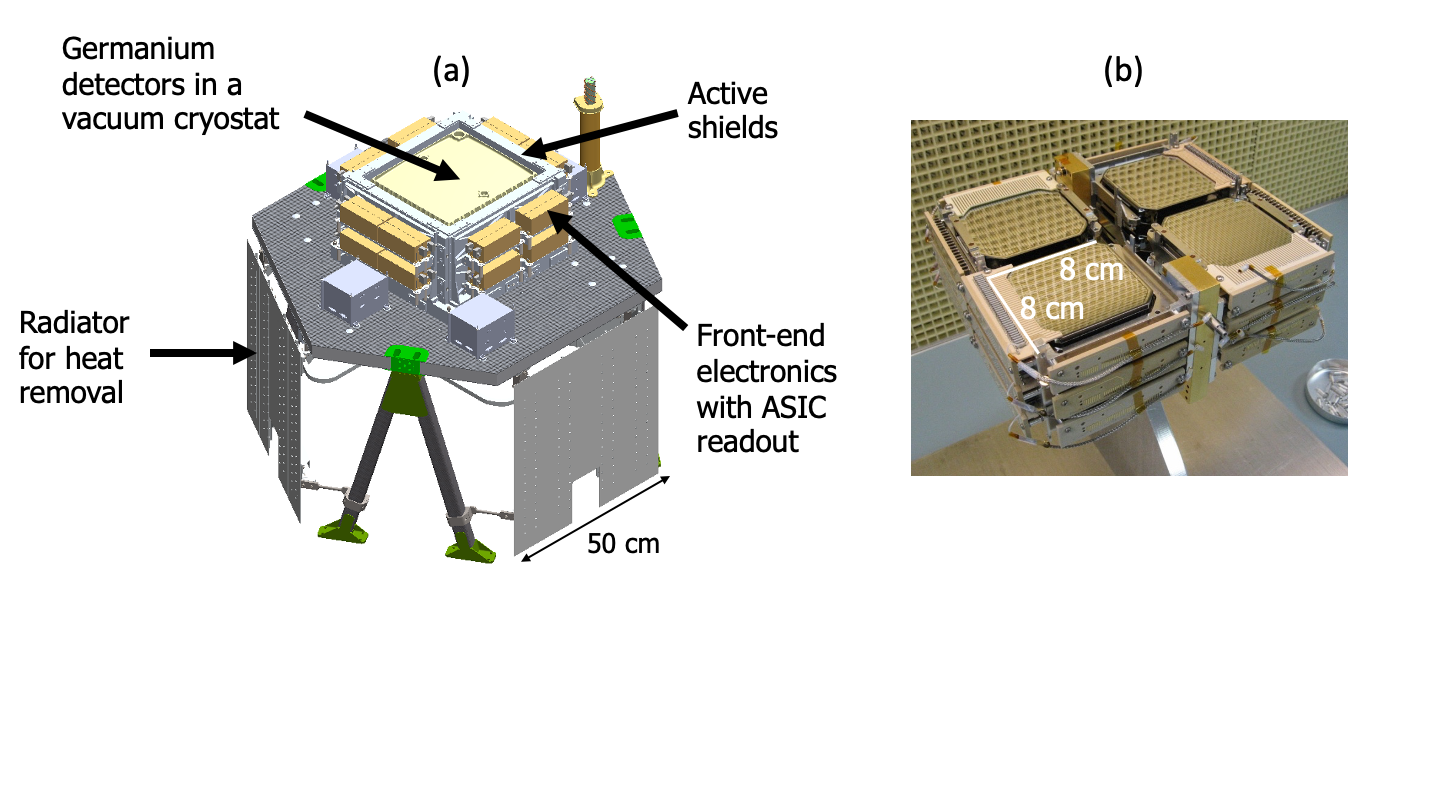}
\end{center}
\vspace{-3.0cm}
\caption{\small (a) The COSI payload design. (b) The 12 germanium detectors (4 stacks of 3) in the COSI-balloon instrument.
\label{fig:cosi_payload}}
\end{figure}

Figure~\ref{fig:cosi_payload}a highlights the main subsystems of the COSI payload:  1.\,the germanium detectors enclosed in a vacuum cryostat; 2.\,the front-end electronics; 3.\,the active shields; and 4.\,the heat removal subsystem.

Semiconducting germanium detectors are used where gamma-ray interactions liberate electrons and holes which, under high-voltage, drift to opposite sides of the detectors.  COSI uses double-sided strip detectors with 64 strips per side and 1.162\,mm strip pitch to measure the $x$ and $y$ interaction positions. The depth ($z$) position is determined by measuring the time difference between the rise of the electron and hole signals.  In addition, the strips are surrounded by a guard ring electrode, which acts as a veto signal.  The detectors are cooled to between 80 and 90\,K using a mechanical cryocooler.  Inside the cryostat, there are 16 detectors (4 stacks of 4), and each detector is 8 $\times$ 8 $\times$ 1.5\,cm$^{3}$.  Figure~\ref{fig:cosi_payload}b shows the 12 detectors (4 stacks of 3) that flew on a balloon platform.  Special processing of the detectors is required \cite{amman20}.

The heart of the front-end electronics is an Application Specific Integrated Circuit (ASIC).  The COSI ASIC has 32 channels with separate circuits for timing, which is required to measure the interaction depth, and spectroscopy \cite{wulf20}.  Each detector side is read out using three ASICs: two for the 64 signal strips and one for the guard ring.  Thus, each detector requires six ASICs, and COSI's 16 detectors use 96 ASICs.

The detectors are surrounded by active shields on four sides and underneath.  The scintillator material used is bismuth germanium oxide (BGO), and they are read out by silicon photomultipliers (SiPMs).  The shields have several functions, including passive background attenuation, active anticoincidence for background radiation coming from the side and bottom, anticoincidence of gamma-rays that escape from the germanium detectors, and finally as a gamma-ray transient monitor.

The largest consumer of power in the satellite is the cryocooler that cools the germanium detectors.  The cryocooler is attached underneath the instrument.  The cryostat heat removal system (CHRS) includes the cryocooler, heat pipes for conducting the heat from the cryocooler, and the radiators.

\section{Measurement requirements for key science goals}
\vspace{-0.3cm}
The key science goals for COSI are: A.\,Uncover the origin of Galactic positrons; B.\,Reveal Galactic Element Formation; C.\,Gain insight into extreme environments with polarization; and D.\,Probe the physics of multimessenger events.  Descriptions of how these goals connect to Astro2020 decadal survey white papers can be found in \cite{tomsick22} and \cite{tomsick19}, and a condensed summary follows.

\begin{figure}[t]
\begin{center}
\includegraphics[height=3.5in,angle=0]{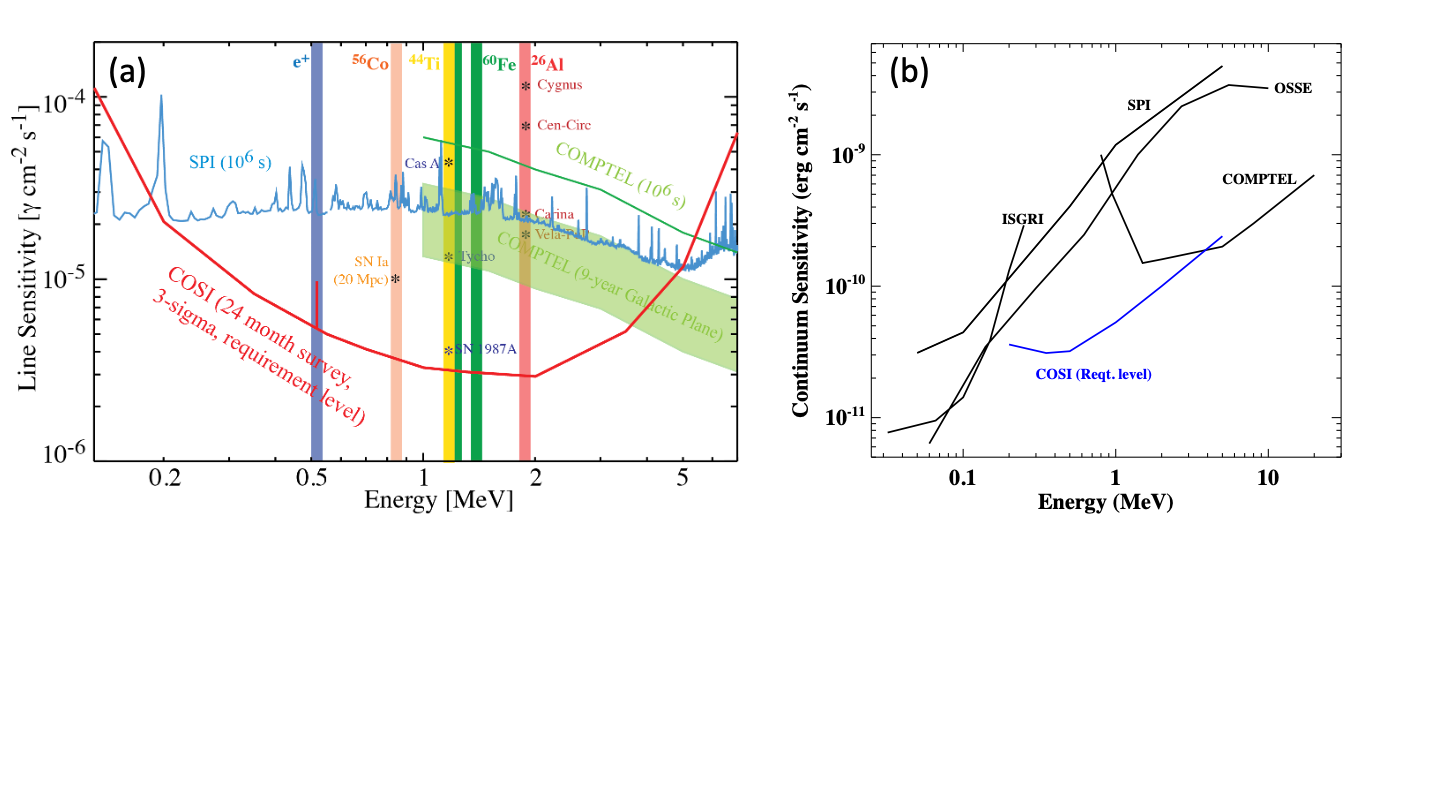}
\end{center}
\vspace{-3.8cm}
\caption{\small Narrow-line (a) and continuum (b) sensitivities based on COSI's requirements compared to current and previous instruments.  The sensitivity curves are for point sources at the 3-$\sigma$ level during 2 years of COSI survey time.  Due to the all-sky coverage that COSI obtains, these sensitivities will be reached for every source in the sky.
\label{fig:sensitivities}}
\end{figure}

The primary measurements for science goals A and B involve making full Galaxy images at 0.511\,MeV and at four nuclear line energies.  The specific spectral resolution, angular resolution, and line sensitivity requirements for making these measurements are provided in Table~\ref{tab:requirements}.  These measurements will allow COSI to determine if the 0.511\,MeV emission from the Galactic bulge has substructure and will measure the latitudinal extension of the Galactic disk.  They will also provide information about the electron-positron annihilation conditions across the Galaxy, including the width of the narrow line and the positronium fraction (by comparing the strength of the narrow line to the orthopositronium continuum).  Measurements at 1.157\,MeV probe $^{44}$Ti from young ($<$1000 years old) supernova remnants.  Measurements at 1.173 and 1.333\,MeV will provide the first Galactic image of $^{60}$Fe, which traces core-collapse supernovae that have occurred over the past millions of years.  Finally, measurements at 1.809\,MeV from $^{26}$Al produced by massive stars will provide advances over past COMPTEL and INTEGRAL maps in terms of sensitivity and angular resolution.  The COSI required line sensitivities are compared to INTEGRAL/SPI and COMPTEL in Figure~\ref{fig:sensitivities}a.

Science goal C requires measurements of polarization, which COSI is sensitive to because the azimuthal component of the Compton scattering angle depends on the polarization of the incident gamma-ray.  The COSI requirement focuses on the measurement of polarization for accreting black holes, and the flux limit given in Table~\ref{tab:requirements} corresponds to being able to measure polarization in the 0.2-0.5\,MeV band for the brightest three known Active Galactic Nuclei (AGN):  Cen~A, 3C~273, and NGC~4151.

\begin{table}[h]
\caption{COSI Requirements\label{tab:requirements}}
\begin{minipage}{\linewidth}
\begin{center}
\scriptsize
\begin{tabular}{ll}
\hline \hline
{\bf Parameter} & {\bf Requirement}\\ \hline
Sky Coverage~~~~~~~~~~~~~~~~~~~~~~~~~~~~~~~~~~~~~~~~~~~~~~~~~~~~~~~~~~~~~~~~~~~~~~~~~~~~~~~~~~~~~~~~~~~~~ & 25\%-sky instantaneous field of view\\ 
 & (all-sky every day in survey mode)\\ \hline
Spectral Resolution (FWHM) & 6.0 keV at 0.511 MeV and 9.0 keV at 1.157 MeV\\ \hline
Angular Resolution (FWHM) & 4.1$^{\circ}$ at 0.511 MeV and 2.1$^{\circ}$ at 1.809 MeV\\ \hline
Line Sensitivity\footnote{3$\sigma$ narrow line point source sensitivity in 2-years of survey observations.} & $1.2\times 10^{-5}$ photons\,cm$^{-2}$\,s$^{-1}$ at 0.511 MeV\\
  & $3.0\times 10^{-6}$ photons\,cm$^{-2}$\,s$^{-1}$ at 1.157 MeV ($^{44}$Ti)\\
  & $3.0\times 10^{-6}$ photons\,cm$^{-2}$\,s$^{-1}$ at 1.173 MeV ($^{60}$Fe)\\
  & $3.0\times 10^{-6}$ photons\,cm$^{-2}$\,s$^{-1}$ at 1.333 MeV ($^{60}$Fe)\\
  & $3.0\times 10^{-6}$ photons\,cm$^{-2}$\,s$^{-1}$ at 1.809 MeV ($^{26}$Al)\\ \hline
Flux limit for polarization\footnote{For 50\% minimum detectable polarization in 2-years of survey observations.} & $1.4\times 10^{-10}$ erg\,cm$^{-2}$\,s$^{-1}$ (0.2-0.5 MeV)\\ \hline
Reporting short GRB detections & $<$1 hour reporting time\\
 & $\sim$1$^{\circ}$ localizations (accuracy depends on GRB fluence)\\
 & 100 ms absolute time accuracy\\ \hline
\end{tabular}
\end{center}
\end{minipage}
\end{table}

While COSI's gamma-ray measurements have connections to all of the other messengers (gravitational waves, neutrinos, and cosmic rays), it is the detection and reporting of short gamma-ray bursts (GRBs) from merging neutron stars that drives COSI requirements related to Goal D.  COSI's large field of view and sensitivity are important for detecting short GRBs.  In addition, the requirement that COSI provides public reports of short GRB localizations within an hour is met by automatically downlinking germanium detector data when gamma-ray transients are detected in the BGO shields.

\section{Extended portfolio of science enabled}
\vspace{-0.3cm}
Beyond the key science goals, COSI's capabilities will enable a large number of additional scientific studies.  We describe a few illustrative examples here, which are taken from an array of investigations planned by the COSI science team. COSI's design is optimized for emission line studies, but it still provides a significant improvement for continuum emission as shown in Figure~\ref{fig:sensitivities}b.  Examples using COSI's emission line and continuum sensitivities are highlighted below.

Blazars, jet-dominated AGN with their jets pointing close to our line of sight, appear to emerge as one of the sources of astrophysical high-energy neutrinos. If these jet environments are efficient neutrino production sites, they are likely to be highly opaque to $\gamma\gamma$ absorption of high-energy and very-high-energy gamma-rays, leading to suppression of the high-energy gamma-ray flux and the initiation of electromagnetic pair cascades. The emission from these cascades is expected to emerge primarily at MeV - sub-MeV X-ray / soft gamma-ray energies \cite[e.g.,][]{gao19,reimer19}.  This makes COSI the ideal instrument to test for possible correlations between very-high-energy neutrinos detected, e.g., by IceCube or KM3NeT, and MeV-flaring blazars, thus providing further evidence for the neutrino - blazar connection. 

Observing gamma-rays at the MeV bandpass also benefits the study of cosmic dark matter (DM). Attractive DM candidates are predicted in the broad mass range of $10^{-22}$\,eV to $10^{35}$\,g, and COSI has the potential to search for many of them. In the ultralight mass region, where the DM mass $M_{\rm DM} << 1$\,eV, the axion-like particle (ALP) is an attractive candidate. COSI will be sensitive to the effects of ALPs on the flux and polarization of MeV gamma-rays emitted from, e.g., blazars \cite{galanti23}. In the light mass region (1\,eV $\lesssim M_{\rm DM} \lesssim $ 100\,GeV), many influential DM candidates have been proposed, including the sterile neutrino, dark photon, and light weakly interacting massive particle (WIMP). Those are predicted around the MeV mass scale, and COSI will observe MeV gamma-ray signals emitted by their decay and annihilation \cite{aramaki_snowmass,caputo23}. Moreover, the candidates also emit low-energy positrons, which are captured by ambient electrons, form positronium, and contribute to 0.511\,MeV emissions. In the heavy mass region where DM is heavier than the EW scale, the heavy WIMP is known to be a well-motivated candidate. Though it primarily emits GeV/TeV gamma-rays by its annihilation, MeV gamma-rays are also produced via inverse Compton scattering caused by energetic electrons from the annihilation. In the ultraheavy mass region where DM is heavier than the Planck scale, primordial black holes (pBH) with mass around $10^{17}$\,g could still compose all of the DM in the present universe. Such pBH would emit both MeV gamma-rays and positrons via Hawking radiation \citep{caputo23}.

The prompt emission spectrum of GRBs is often described by the Band function. But the underlying radiation mechanisms are still debated. Synchrotron emission in either ordered or random magnetic fields can contribute to the prompt emission, while photospheric emission may also explain the observed spectrum. COSI can distinguish these mechanisms by measuring the MeV polarization, since they produce distinct polarization signatures \cite{toma09,wei16,gg21,pl22}: synchrotron polarization degree depends on the orderness of the magnetic field, while the photospheric emission may produce polarization degree from 0 to 100\% depending on the geometry of the Compton scattering, which is probably anisotropic in the prompt emission region. By observing the polarization degree of $>$30 GRBs, COSI will be able to tell synchrotron in an ordered magnetic field apart from synchrotron in a random magnetic field and photospheric models, since the latter two are characterized by a different number distribution of GRB prompt polarization. Additionally, COSI may observe the time-dependent polarization signatures for a few bright GRBs during its mission span, which can further probe the underlying physical conditions in those bright GRBs.

Thermonuclear supernovae (SN Ia) are important tools of modern cosmology and significant contributors to galactic nucleosynthesis and energetics, yet their progenitor systems and explosion mechanisms are poorly understood. As shown in Figure~\ref{fig:sensitivities}a, the expected peak flux of the prominent 0.847\,MeV line of $^{56}$Co decay is below the 10$^{6}$\,s COSI sensitivity for a SN Ia at distance 20\,Mpc. However, with COSI's complete coverage of the $\sim$10$^{7}$\,s light curve peak and sensitivity to other lines, e.g., 1.238\,MeV, COSI should detect $^{56}$Co from SN Ia at this distance, within which there is roughly one per year. For nearer SN Ia, like the previous SN 2011fe and SN 2014J (see, e.g., \cite{churazov15}), COSI could monitor the line light curves and resolve their velocity profiles, yielding independent determinations of the synthesized $^{56}$Ni mass, its distribution within the ejecta, and total ejecta mass and kinetic energy \cite{leising22}. These would provide important constraints on the thermonuclear explosions and white dwarf progenitor systems. 

Classical nova explosions are caused by thermonuclear runaways of material accreted by white dwarfs in a close binary systems. The material ejected at a velocity of several thousands of km/s is enriched with freshly synthesized nuclei. Among these, the unstable $^{13}$N and $^{18}$F decay with a lifetimes of 10\,min and 1.8\,hr, respectively, by emitting positrons that rapidly annihilate in the dense ejecta in expansion. The resulting annihilation photons and their Comptonization produce a prompt emission ($\sim$2 hr) of a 0.511\,MeV line and a gamma-ray continuum $<$0.511\,MeV \citep{gomezgomar98, hernanz14}, which has not been detected yet \citep{senziani08, siegert18}. The long-lived radioactive nuclei synthesized in the explosion are $^{7}$Be in CO white dwarfs and $^{22}$Na in ONe white dwarfs. Their lifetimes (53\,days and 2.6\,yr, respectively) are long enough for them to decay in a mostly transparent ejecta by emitting gamma-ray lines at 0.478\,MeV and 1.275\,MeV, respectively.  The detection of the prompt emission with COSI would provide information on the outburst conditions and on the properties of the ejecta, while the detection of 0.478\,MeV and 1.275\,MeV lines would provide the type (CO or ONe) of the underlying white dwarf and the mass of released radioactive isotopes, which give unique constraints to explosive nucleosynthesis models. 

COSI's continuum sensitivity will also provide new constraints on MeV-band emission from the Fermi Bubbles \citep{negro22} and the Galactic Diffuse Continuum Emission (GDCE).  The GDCE in the COSI energy band is dominated by inverse Compton radiation, produced by cosmic ray (CR) electrons up-scattering low-energy photons of the interstellar radiation field. There is also expected to be subdominant contributions from Bremsstrahlung radiation, unresolved point sources, and plausibly other sources as well. COSI's continuum sensitivity and wide field of view will enable measurements of a highly precise spectrum of the GDCE over the entire sky. Such measurements will play a major role in disentangling the different components of the emission, including searches for new physics, and they will likely lead to new insights regarding the sources of CR electrons in the Galaxy. More details regarding COSI's prospects for probing the GDCE can be found in the ICRC2023 proceedings paper by Karwin et al. 

\vspace{1cm}
\noindent
{\bf \large Acknowledgements:} COSI is a NASA Small Explorer mission and is supported under NASA contract 80GSFC21C0059. The research of Dr. Lewis, Dr. Zhang, and Dr. Karwin was supported by an appointment to the NASA Postdoctoral Program at the NASA Goddard Space Flight Center, administered by Oak Ridge Associated Universities under contract with NASA.


%
%
%

\end{document}